\documentstyle[12pt]{article}

\def\frac#1#2{{\textstyle {#1 \over #2}}}

\def\Eq{\begin{equation}}   \def\Endeq#1{\label{#1} \end{equation}}
\def\Eqa{\begin{eqnarray}}  \def\Endeqa#1{\label{#1} \end{eqnarray}}

\newcommand{\Eqr}[1]{(\ref{#1})}

\def\gev{\rm\,GeV}

\def\&{and}
\def\bps{ \bar{\psi} }
\def\pbp{ \langle \bps \psi \rangle }
\def\xsb{ \chi S B }
\def\Q {Q^i}
\def\Qt {\tilde{Q}^i}
\def\DS {D\!\!\!\!/}

\def\ap#1#2#3{           {\it Ann. Phys. (NY) }{\bf #1}, #2 (19#3)}

\def\np#1#2#3{           {\it Nucl. Phys. }{\bf #1}, #2 (19#3)}
\def\pl#1#2#3{           {\it Phys. Lett. }{\bf #1}, #2 (19#3)}
\def\pr#1#2#3{           {\it Phys. Rev. }{\bf #1}, #2 (19#3)}

\def\prl#1#2#3{          {\it Phys. Rev. Lett. }{\bf #1}, #2 (19#3)}

\begin{document}
\begin{titlepage}

\begin{center}
January, 1993      \hfill       HUTP-92-A047\\
\vskip .5 in
{\large \bf Complementarity and Chiral Fermions in SU(2) Gauge Theories}
\vskip .3 in
{
  {\bf Stephen D.H. Hsu}\footnote{Junior Fellow, Harvard Society of
     Fellows. Email: \tt Hsu@HUHEPL.bitnet, Hsu@HSUNEXT.Harvard.edu}
   \vskip 0.3 cm
   {\it Lyman Laboratory of Physics,
        Harvard University,
        Cambridge, MA 02138}\\ }
  \vskip 0.3 cm
\end{center}

\vskip .5 in
\begin{abstract}
Complementarity - the absence of a phase boundary separating the Higgs
and confinement phases of a gauge theory - can be violated by the
addition of chiral fermions. We utilize chiral symmetry violating
fermion correlators such as $ \langle \bps \psi \rangle $
as order parameters
to investigate this issue. Using inequalities similar to those of
Vafa-Witten and Weingarten, we show that SU(2) gauge theories with
Higgs and fermion fields in the fundamental representation exhibit
chiral symmetry breaking in the confined phase and therefore do {\it not}
lead to massless composite fermions. We discuss the implications for the
Abbott-Farhi strongly interacting standard model.

\end{abstract}
\end{titlepage}

\renewcommand{\thepage}{\arabic{page}}
\setcounter{page}{1}
\section{Complementarity and all that}
Certain gauge theories with scalars in the fundamental representation
can be shown to exhibit a remarkable property known as complementarity
\cite{OS,FS,DRS}.
Complementarity means that the Higgs phase (large vacuum expectation value v,
small gauge coupling g) and confinement phase (small v, large g) are not
separated by a phase boundary. (Here both $g = g( \Lambda )$ and v are
defined in terms of some lattice spacing $\Lambda^{-1}$.) The result,
proved by Fradkin and Shenker \cite{FS} using results of Osterwalder
and Seiler \cite{OS} (see also Banks and Rabinovici \cite{BR} for a
similar result for U(1) theories), consists of demonstrating that in
a lattice formulation of the theory all correlators (ie free energy,
n-point Greens functions) are analytic functions of g and v in a connected
region which contains both the Higgs and confinement phases.
(See figure 1 for a typical phase diagram.)
Therefore, quantities such as the free energy of the theory vary
smoothly without discontinuity as we interpolate between the two regions.

The rigorous demonstration of complementarity coincided with observations
by 't Hooft \cite{TH} and Susskind (unpublished) that there exists a strong
similarity between the spectrum of states in the standard Higgs picture and
the confined picture of a gauge theory with scalars in the fundamental.
This led 't Hooft to remark that the question of confinement in this
class of models could only be answered dynamically - there being no
fundamental difference between the confining and spontaneously broken phases.
(Indeed, these remarks apply equally to QCD, despite its lack of fundamental
colored scalars, because of composite fields which can be formed out of glue
and fermions.)

In this letter we wish to examine complementarity
in gauge-Higgs models when chiral fermions are included.
We will demonstrate in the
SU(2) case that the addition of chiral fermions is capable of drastically
altering the phase diagram of the theory. The models we study exhibit a
phase transition associated with chiral symmetry breaking ($\xsb$) as
we move from the broken to confined phase.
The above result was established previously by I.-H. Lee and R. E. Shrock
\cite{Shrock} using analytical and numerical techniques on the lattice.
Our analysis will be in the continuum, which makes it less rigorous
from the viewpoint of constructive quantum field theory but perhaps easier
to understand to theorists who work in the continuum.

It is straightforward to argue that addition of chiral fermions to a purely
bosonic theory can lead to a violation of complementarity. One has
merely to consider the 't Hooft anomaly matching conditions \cite{TH},
which are necessary but not sufficient conditions for the existence of
massless composite fermions.
If the anomalies resulting from the fundamental fermion triangle
graphs do not match those of the (putative) massless composite
fermions, one can immediately conclude that there are no massless
composites and chiral symmetries are broken. The effects of the
anomaly are then reproduced by the Goldstone modes via the Wess-Zumino
term \cite{WZ} in the chiral lagrangian. Here we will examine a more
interesting class of models, where the 't Hooft matching conditions
{\it are} satisfied by the composite fermions, and therefore dynamical
information is necessary to determine the status of the chiral symmetries.
There has been much interest in models of this type in which the known
fermions of the standard model are massless bound states of more fundamental
preons \cite{Preon}. Dimopoulos, Raby and Susskind \cite{DRS} gave a
physically motivated construction involving tumbling (via fermionic
condensates) of chiral gauge theories which always yields solutions
to the 't Hooft conditions.

More information is required to conclude
that models of the above sort actually yield
massless composites. The only cases
this author is aware of where the existence of massless composites
can be demonstrated is in models where the large-N approximation (see
Eichten et. al. \cite{EPPZ}) can be used to show the absence of goldstone
bosons at leading order in $1/N$. Here we will prove that in a certain
class of SU(2) gauge models, which satisfy both the bosonic complementarity
conditions and 't Hooft's matching conditions, chiral symmetry is indeed
broken and no massless composites are formed. The essence of the argument
is that since representations of SU(2) are real, the model can be rewritten
as a vectorlike gauge theory. Weingarten \cite{W} and Vafa and Witten
\cite{VW} have shown that in vectorlike theories
rigorous inequalities apply to the correlators of
certain conserved currents. Using a similar analysis, we show
that the mass of a composite fermion formed of a scalar and a
fundamental fermion is nonzero in the chiral limit of the model.
This is sufficient to rule out the massless composite realization of
't Hooft's conditions and therefore implies the existence of Goldstone
bosons and $\xsb$.
The above result stands in contradiction to one of the key dynamical
assumptions of
the Abbott-Farhi strongly interacting standard model, which
relies on the existence of (nearly) massless composites. In
the following section we will present our argument for $\xsb$
in certain SU(2) theories, and discuss the implications for the
phase diagrams of those theories. In the final section we will
discuss the implications for the Abbott-Farhi model and preonic models.

\renewcommand{\thepage}{\arabic{page}}
\section{Fermions and chiral symmetries}

Consider an SU(N) gauge theory with (N-1) Higgs bosons in the fundamental
representation (sufficient to completely break the gauge symmetry)
\footnote{For the lattice proofs \cite{OS,FS} to apply it is necessary that
the gauge symmetry be completely broken by the Higgs fields. It is often
incorrectly stated in the literature that having the Higgs fields in
the fundamental representation is sufficient for complementarity.}.
It can be rigorously shown using lattice methods \cite{OS,FS} that this
theory exhibits complementarity and exhibits a phase diagram similar to
the one in figure 1.
Now consider adding $N_f$ massless, chiral fermions $\psi_i$ to the theory.
At the classical level there is an exact $SU(N_f)$
chiral symmetry associated with the $N_f$ fermions.
In what follows we will use chiral symmetry violating correlators such as
$\pbp \equiv \langle \bps_i^{\alpha} \psi_{i \alpha} \rangle$
as order parameters to investigate the phase diagram for this theory.
Here $\alpha$ is an SU(N) index and the flavor index i is left arbitrary.
A nonzero value of $\pbp$ for any value of i will be sufficient to show a
violation of complementarity.

In the perturbative Higgs regime (g small, v large)
these symmetries remain unbroken by quantum effects.
We can argue this result as follows: suppose the weak
coupling effects are sufficient to break some of the
chiral symmetries. Then by the Goldstone theorem there
must exist massless composite Goldstone bosons, formed
from the massless fermions. However the binding energy of
the composite must be sufficient to cancel the positive
kinetic energy of the two fermions confined to a region
of the size of the Goldstone boson. Since in weak coupling
one expects the binding energy to be proportional to the
fermion mass $m_f$, this is impossible if the size of the
Goldstone boson is to remain finite in the zero mass limit.
(A more rigorous, lattice argument for the absence of spontaneous
symmetry breaking at arbitrarily weak coupling has been given by
Lee and Shrock. See the early papers in \cite{Shrock}.)

The above argument holds for sufficiently small coupling g.
Therefore there must exist a small patch in the upper left
hand corner of figure 1, in which there is no $\xsb$ and
the order parameter $\pbp$ is exactly zero. However, by
analyticity\footnote{Technically, to apply analyticity
we must verify that the correlator still vanishes when
the parameters g and v are given infinitessimal imaginary
parts. We will assume this is the case. The lattice proofs
of analyticity of course still apply when g and v have small
imaginary parts.}, the vanishing of this correlator can be
extended throughout the entire region where complementarity applies.
In particular, if figure 1 truly represents the phase diagram
of the theory we can conclude that $\xsb$ does not occur in
the confined phase. If, on the other hand, we can demonstrate
that chiral symmetries are broken in the confined phase it will
imply the existence of a phase boundary between the Higgs and
confinement regions, and a violation of complementarity.

We will now proceed to show that chiral symmetries are
indeed broken in the confined phase of the above theory
when $N = 2$. (We continue to assume as above that there
are no Yukawa couplings between the scalar and fermions
and no explicit $\xsb$.) This is a very plausible result
for SU(2) gauge theories for the following reason: because
representations of SU(2) are real, it is always possible to rewrite a
$N_f$ flavor SU(2) theory as a {\it vectorlike} $N_f/2$ flavor theory.
($N_f$ must be even to guarantee vanishing of Witten's global
SU(2) anomaly \cite{WGA}. )
The latter theory, in the absence of
fundamental scalars, is merely two color
QCD which certainly breaks its chiral symmetries.
Therefore, unless the presence of fundamental scalars
somehow dramatically alters the dynamics of the theory,
we expect the same to be true here.
In particular, it is clear that if the scalar mass is
taken to infinity, thereby decoupling it from the low
energy dynamics, $\xsb$ must occur.

We will show that $\xsb$ occurs for a large range of values
of the scalar mass.
Our strategy is to prove that composites with interpolating fields
given by
\begin{eqnarray}
Q^i &\equiv& \phi^{* \alpha} \psi_{\alpha}^i \\
\tilde{Q}^i &\equiv& \phi_{\alpha} \epsilon^{\alpha \beta} \psi^i_{\beta}
\end{eqnarray}
are not massless as long as the scalar mass is
nonzero\footnote{In other words, we will prove
that the $N=2$ theories satisfy a ``persistence of mass''
condition \cite{PW}. }.
In the confined phase of the theory there are only two
candidates for matching the anomalies of the fundamental fermions $\psi^i$:
the SU(2) singlet composites $\Q, \Qt$ and the Goldstone bosons
$\pi^{ij} \sim \bps^i \psi^j$. If $\Q,\Qt$ are not massless, the
matching conditions must be satisfied by the Goldstone bosons and therefore
$\xsb$ must occur.

We first rewrite the model in a vectorlike manner. Define
\begin{equation}
\chi^i ~=~ (\psi ^{i + N_f/2})^c ~=~ i \gamma_2 (\psi ^{i + N_f/2})^*.
\end{equation}
Note that the $\chi$ fields have the opposite
chirality of the $\psi$ fields, but are still doublets.
The theory is clearly vectorlike as we can now add gauge
invariant mass terms to the Lagrangian, pairing $\chi^i$ with $\psi^i$.

Now let $\Psi^i = \psi^i + \chi^i$. With the addition to the Lagrangian
of mass terms $m_i \bar{\Psi}^i \Psi^i$ our theory is now simply $N_f/2$
flavor, two color QCD with massive dirac fermions and an extra colored scalar.
The fermion masses and the ``vectorization'' of the model are necessary
for technical reasons in order to apply certain rigorous results similar
to those first derived by Vafa and Witten \cite{VW}. At the end of the
calculation we will take $m_i \rightarrow 0$ to reduce it to the original,
with classical chiral symmetries intact.

Let us temporarily redefine the interpolating
fields $\Q,\Qt$ so that they each contain a
Dirac fermion $\Psi^i$ rather than a Weyl fermion as
previously defined. The index i now runs from 1 to $N_f/2$.
If massless composite fermions are to exist in the limit
$m_i \rightarrow 0$ corresponding to the old $\Q,\Qt$, then
the new $\Q,\Qt$ must also be massless in that limit. We will
now demonstrate that this is not the case.

Consider the Euclidean propagator for the
Q field: (From here on we selectively suppress
flavor and color indices for simplicity. Q refers to either of $\Q,\Qt$.)
\begin{equation}
\langle T(\bar{Q}(x) Q(y)) \rangle = Z^{-1}
\int DA~ D\phi~ D\bar{\Psi}~ D\Psi~  exp( - S_E[A, \phi, \Psi] )~~
\bar{Q} (x) Q (y),
\end{equation}
where $Z^{-1}$  is the standard normalization factor and the Euclidean action
is
\begin{equation}
S_E[ A, \phi, \Psi] = \int d^4x~\frac{1}{2g^2}TrFF + |D \phi|^2 + M^2 \phi^2
             + \sum_i \bar{\Psi} ( \DS + m_i  ) \Psi.
\end{equation}
Figure 2 gives a pictorial description of the expectation value in equation 4.
The normalization factor $Z^{-1}$  divides out all vacuum bubbles, so we are
left with $\phi$ and $\Psi$ propagators summed over all possible gauge
backgounds, with scalar and fermion loops included. Conservation of
flavor and scalar number (for the moment we neglect scalar self-interactions)
prevents either the $\Psi$ or $\phi$
lines from terminating except on an insertion of Q.

We can rewrite (4) in the following manner by integrating
out the scalar and fermion fields:
\begin{eqnarray}
\langle T(\bar{Q}(x) Q(y)) \rangle
&=& Z^{-1} \int D\mu~ (D^2+M^2)^{-1}_{A,xy}~(\DS + m)^{-1}_{A,xy}     \\
D\mu &\equiv&  DA~exp(-\int d^4x~\frac{1}{2g^2}TrFF)~
{\det}^{-1/2}(D^2 + M^2) \prod_i {\det}
 (\DS + m_i).
\end{eqnarray}
Here the determinants and propagators are evaluated in an arbitrary
gauge backgound $A_{\mu}$, which is then integrated over. The key
point is that the measure of integration $D\mu$ can be shown to be
positive definite. This is because both the scalar and gauge field
Euclidean actions are real and the fermion determinant is always
real and positive in a vectorlike theory \cite{VW}. (It is also
necessary to choose the topological $\theta$ term to be zero.)

Since the measure is positive definite, any $A_{\mu}$ independent bound
that can be placed on the integrand will yield a bound on the Q propagator.
The above integrand consists of the product of the scalar and fermion
propagators in arbitrary gauge background. A great deal is known about the
behavior of such propagators at large separations $|x-y|$. For example,
Kato's inequality \cite{Kato} asserts that
$| (D^2+M^2)^{-1}_{A,xy} | \leq | (D^2+M^2)^{-1}_{A=0,xy} |$.
That is, the scalar propagator in an arbitrary gauge background
falls off faster than its free counterpart (ie in zero gauge field background).
A similar result, involving for technical reasons a {\it smeared}
fermion propagator also applies.
Here we will sketch the arguments
from \cite{VW} which apply to a smeared propagator of
either scalar or fermionic type. Consider the smeared propagator
\begin{equation}
(\DS + m)^{-1}_{A,\alpha \beta} \equiv \langle \alpha | (\DS + m)^{-1}_{A}
                | \beta \rangle ,
\end{equation}
where $| \alpha \rangle,| \beta \rangle$ are localized
wave packet states, rather than position eigenstates
$|x \rangle, |y \rangle$. By a wave packet state we mean that
$\Psi(x) | \alpha \rangle = \phi(x) | \alpha \rangle = 0$ outside
a compact region $\alpha$ centered at x with size much smaller than $|x-y|$.

We can bound the smeared propagator by the following trick:
\begin{eqnarray}
\langle \alpha | (\DS + m)^{-1}_{A} | \beta \rangle
      &=& \int_0^{\infty} dt~ \langle \alpha |
           exp[- (\DS + m)_{A} t ] | \beta \rangle \\
       &=& \int_0^{\infty} dt~ e^{-mt}~  \langle \alpha |
           exp[- i ( -i \DS) t ] | \beta \rangle.
\end{eqnarray}
The last expression has the form of a quantum mechanical
transition amplitude in a (4+1) dimensional theory with
Dirac Hamiltonian $H = -i \DS$. By causality, we have
$\langle \alpha | e^{-iHt} | \beta \rangle = 0$ for
$0 \leq t < |x-y|$. Therefore
\begin{eqnarray}
\langle \alpha | (\DS + m)^{-1}_{A} | \beta \rangle &=&
       \int_{t= |x-y|}^{\infty} dt~ e^{-mt}~ \langle
\alpha | e^{-iHt} | \beta \rangle , \\
|\langle \alpha | (\DS + m)^{-1}_{A} | \beta \rangle|
      &\leq& 1/m~ e^{- m|x-y|} | \alpha | | \beta | ,
\end{eqnarray}
where we have used (4+1) unitarity and the Schwarz inequality
to obtain the last expression, and
$| \alpha |, | \beta |$ are the norms of the states
$ | \alpha \rangle,| \beta \rangle$. A similar result
applies in the scalar case. Note that for position
eigenstates the corresponding norms are infinite and
hence do not yield a useful bound.

Putting the above results together, we have:
\begin{equation}
| \langle T(\bar{Q}(x) Q(y))_s \rangle | \leq \frac{C}{m} exp( - (M+m)|x-y| ),
\Endeq{bound}
where the subscript ``s'' means smeared and C is a numerical constant.
This bound precludes the existence of a massless bound state Q in the limit
of zero fermion mass and massive scalar ($m \rightarrow 0, M$ fixed).
Note that this type of bound does {\it not} preclude the existence of a
massless bound state consisting of {\it two} massless fermions, as in
that case as $m \rightarrow 0$ the bound disappears. This is crucial, as
we expect to find massless composite Goldstone bosons
$\pi^{ij} \sim \bps^i \psi^j$ in the chiral limit.

An important technical point is that the above arguments require a
cutoff, $\Lambda$, for the theory because the masses $M,m$ appearing
in the various inequalities are actually bare masses, $M_0,m_0$. We
can either imagine that this entire analysis has been conducted on
the lattice (see \cite{W}), or that a suitable, gauge invariant
regularization such as Pauli-Villars has been carried out. Because
of this technical requirement, there is a problem with the inequality
(13) which stems from the unnaturalness of scalar models. The problem
 is that as the cutoff $\Lambda$ is taken to $\infty$, the {\it bare}
scalar mass required to yield a fixed {\it physical} scalar mass becomes
negative. This is easy to see in perturbation theory, as the one loop
correction to the bare mass has the form
\begin{equation}
M^2_{physical} = M^2_0 + \frac{\lambda}{32 \pi^2} \Lambda^2
+ \frac{3 g^2}{32 \pi^2} \Lambda^2,
\Endeq{quad}
where $\lambda, g$ are respectively the $\phi^4$ and gauge couplings and
we have computed the latter in Landau gauge. We have suppressed subleading
logarithmic corrections. Note that if one wishes to take the cutoff
arbitrarily large with respect to the physical scalar mass, an
arbitrarily large and negative bare mass is required. Therefore,
as $\Lambda \rightarrow \infty$ our bound \Eqr{bound} becomes useless.

It is possible to choose bare parameters such that (13) holds with
positive $M_0^2$ if the cutoff is not chosen too large.
(We also require $\lambda(\Lambda) = 0$ in order to perform the
scalar functional integration in Eq. (4) \footnote{This choice of
$\lambda(\Lambda)$ does not imply the theory is unbounded from below. If one
computes the renormalization group improved effective potential, it is easy to
see that a positive renormalized mass squared compensates for $\lambda(\mu)$
being driven (logarithmically) negative in the infrared. There is also no
vacuum expectation induced for the scalar unless the ratio $\Lambda /
M_{physical}$ is taken very large.}.)
One would
like to keep the cutoff large compared to the scale $\Lambda_2$ at
which $SU(2)$ becomes strongly interacting, while keeping
$M_{physical} \simeq \Lambda_2$. Whether this is possible
depends on the evolution of the gauge coupling constant
between $\Lambda$ and $\Lambda_2$. This in turn depends
on the number of fermion flavors in the model. If we fix
$M_{physical} = \Lambda_2$, for $N_f = 2$ we get
$\Lambda / \Lambda_2 \simeq 3.5$ while for $N_f = 12$
(the Abbott-Farhi case) we get $\Lambda / \Lambda_2 \simeq 2$.
(Note that for $N_f = 2$ the axial chiral symmetry is anomalous,
and therefore already explicitly violated by quantum effects.)
Larger ratios of $\Lambda / \Lambda_2$ are possible if we allow
$M_{physical} > \Lambda_2$. This verifies our intuition that very
heavy scalars should
decouple from the strong $SU(2)$ dynamics,
leaving behind a theory with broken chiral symmetries. However, the
Abbott-Farhi model assumes a phase transition between $\xsb$ and
no $\xsb$ as the mass of the scalar is lowered, and hence we are
more interested in what happens when
$M_{physical} \simeq \Lambda_2$.

Should it concern us that the cutoff must be taken so low?
If we think of the renormalization group in Wilson's language we know that we
can start with a continuum theory (or one with arbitrarily large cutoff,
momentarily ignoring triviality problems of scalar theories) and relate it
to an effective theory at scale $\mu$ by systematically integrating out
degrees of freedom. The information from the high momentum modes will be
contained in the running coupling constants $ g_i (\mu)$ and the higher
dimension operators $O_i (\mu)$ induced by this procedure.
The theory defined with cutoff $\Lambda$, ``bare'' couplings equal
to $g_i (\Lambda)$ and additional non-renormalizable interactions given by
$O_i(\Lambda)$ is then completely equivalent to the original continuum theory.

In deriving our inequalities we assumed that there were no non-renormalizable
operators in the bare theory. Perhaps this is justified because
we are attempting to determine whether an exactly massless state exists in a
certain channel. We are therefore interested in physics at very low momentum
scales, and at arbitrarily long distances, where higher-dimension operators
should be irrelevant. One might expect that the long distance behavior of two
theories differing by some set of such irrelevant
operators $O_i$ should be the same.
(By assumption the operators $O_i$ respect the fermion chiral symmetries,
otherwise complimentarity is already violated.)
However, a loophole in this line of reasoning is that the higher-dimension
operators may actually shift the ground state of the theory from chiral
symmetry preserving to breaking. This is possible in principle if
spontaneous $\xsb$ is
due only to gauge dynamics of momentum scale $k \simeq \Lambda_2$ rather
than $k << \Lambda_2$. In that case the operators $O_i$ are only supressed
by powers of $\Lambda_2 / \Lambda$ to some power, and may be important if this
ratio is not large.
Because of this possibility, our results can probably only be
rigorously applied for $M_{physical} > (few)~ \Lambda_2$.
This leaves a small window for the Abbott-Farhi model with a very light Higgs
scalar. However, the lattice work of Lee and Shrock \cite{Shrock} is
valid for all values of the scalar mass, and hence is sufficient to close this
window.

\renewcommand{\thepage}{\arabic{page}}
\section{Further implications: the Abbott-Farhi model}

The existence of a chiral phase boundary in the above class of models
has strong implications for the Abbott-Farhi model.
Abbott and Farhi \cite{AF} formulated a strongly coupled version of the
electroweak theory, in which the the Higgs vacuum expectation value is small
and the SU(2) coupling large. This model is successful in roughly reproducing
the observed spectrum of the electroweak theory, generating W and Z bosons as
bound states of the scalar doublet field, although its current
phenomenological viability is subject to certain unproven dynamical
assumptions (see last reference in \cite{AF}).

One of the key dynamical assumptions is that chiral symmetries remain unbroken
in the confined phase and that in the limit of zero fermion-Higgs Yukawa
couplings there are massless, left handed fermionic bound states $Q^i_L$.
Nonzero Yukawa couplings $\lambda_i$ provide mass terms
which marry the $Q^i_L$ to SU(2)
singlet right handed fermions $\psi^j_R$ yielding Dirac fermions of mass
$\sim \lambda_i \Lambda_{SU(2)}$, where $\Lambda_{SU(2)} \simeq 256 \gev$ is
approximately the weak scale.

Since the electroweak theory is intrinsically chiral, one might wonder
how the results of the previous section can be applicable. The answer is that,
perhaps suprisingly, the electroweak theory is actually vectorlike in the
limit where we ignore hypercharge and color.
In the absence of hypercharge, color and Yukawa
couplings the electroweak theory belongs to the class of models studied in
section 2. (It is easy to show that the electroweak theory has no physical
$\theta$ angle.) Therefore, to the extent to which those couplings can be
treated as perturbations the results of the previous section should apply to
the Abbott-Farhi model.
Indeed, all of the relevant coupling constants are small at scale $\Lambda_2$,
 with the possible exception of the top Yukawa coupling.
This suggests that the masses of the fermions in that
theory do {\it not} resemble those of their perturbative electroweak
counterparts. We expect the right handed fermions to remain nearly
massless, with masses of order $(\lambda v)^2 / M_L$ induced by their
interactions
with the $Q^i_L$.  We also expect a plethora of relatively light
 pseudo-Goldstone
bosons which are bound states of left handed leptons and quarks.
Finally, the condensates which form will spontaneously break color and
hypercharge.

Because our results are specific to gauge groups which have (pseudo)real
representations, it is not always possible to apply them to more complicated
composite models \cite{Preon}. However, it seems possible to this author that
many models of the sort first constructed by Dimopoulos et. al. \cite{DRS} may
indeed undergo $\xsb$ rather than produce massless composite fermions in the
confined phase. Additional dynamical information is required to decide the
issue.

\section{Acknowledgments}

The author would like to thank
E. Farhi, H. Georgi, T. Gould,
V. Khoze, S. Nussinov and R. Singleton for useful discussions.
The author is also grateful to R.E. Shrock for discussions of
his lattice work in this context.
SDH acknowledges
support from the National Science Foundation under grant
NSF-PHY-87-14654,
the state of Texas
under grant TNRLC-RGFY106, the Milton Fund of the
Harvard Medical School and from the Harvard Society of Fellows.

\vskip 1 in
{\bf figure 1:} Typical phase diagram for theories exhibiting
complementarity. All physical quantities are analytic functions
of v and g in the shaded region.
\vskip .3 in
\noindent {\bf figure 2:} Q field propagator in arbitrary
gauge field background.
\end{document}